\documentclass[10pt]{article}

\usepackage{amsmath}
\usepackage{amssymb}

\usepackage{graphicx}
\usepackage{epstopdf}
\usepackage{cite}

\usepackage{color} 

\makeatletter
\renewcommand{\@biblabel}[1]{\quad#1.}
\makeatother

\date{}

\begin{document}

% Title must be 150 characters or less
\begin{flushleft}
{\Large
\textbf{Epidemiological modeling of online social network dynamics}
}
% Insert Author names, affiliations and corresponding author email.
\\
John Cannarella$^{1}$, 
Joshua A. Spechler$^{1,\ast}$
\\
\bf{1} Department of Mechanical and Aerospace Engineering, Princeton University, Princeton, NJ, USA
\\
$\ast$ E-mail: Corresponding spechler@princeton.edu
\end{flushleft}

% Please keep the abstract between 250 and 300 words
\section*{Abstract}
The last decade has seen the rise of immense online social networks (OSNs) such as MySpace and Facebook. In this paper we use epidemiological models to explain user adoption and abandonment of OSNs, where adoption is analogous to infection and abandonment is analogous to recovery. We modify the traditional SIR model of disease spread by incorporating infectious recovery dynamics such that contact between a recovered and infected member of the population is required for recovery. The proposed infectious recovery SIR model (irSIR model) is validated using publicly available Google search query data for ``MySpace" as a case study of an OSN that has exhibited both adoption and abandonment phases. The irSIR model is then applied to search query data for ``Facebook," which is just beginning to show the onset of an abandonment phase. Extrapolating the best fit model into the future predicts a rapid decline in Facebook activity in the next few years.

% Please keep the Author Summary between 150 and 200 words
% Use first person. PLoS ONE authors please skip this step. 
% Author Summary not valid for PLoS ONE submissions.   
%\section*{Author Summary}

\section*{Introduction}

Online social networks (OSNs) have revolutionized interpersonal interaction by greatly facilitating communication between users, leading to the popularity of OSNs seen today \cite{Boyd2007}. The rise to ubiquity of OSNs has been accompanied by the creation of a multibillion dollar industry to provide these social networking services. Two high profile examples are Facebook (\$140 billion) and Twitter (\$35 billion) \cite{googlefinance}, both of which command high valuations based on their large user bases and expectations of growth. Despite the recent success of Facebook and Twitter, the last decade also provides numerous examples of OSNs that have risen and fallen in popularity, most notably MySpace. MySpace, founded in 2003, reached its peak in 2008 with 75.9 million unique monthly visits in the US before subsequently decaying to obscurity by 2011\cite{myspaceFall}. MySpace was purchased by News Corp for \$580 million during its rapid growth phase in 2005 and was sold six years later at a loss for \$35 million in 2011 \cite{GrantMeadows2012}. The dynamics governing the rapid rises and falls of OSNs are therefore not only of academic interest, but also of financial interest to incumbent and emerging OSN providers and their stakeholders.

In this paper we analyze the adoption and abandonment dynamics of OSNs by drawing analogy to the dynamics that govern the spread of infectious disease. The application of disease-like dynamics to OSN adoption follows intuitively, since users typically join OSNs because their friends have already joined \cite{Wu2013}. The precedent for applying epidemiological models to non-disease applications has previously been set by research focused on modeling the spread of less-tangible applications such as ideas\cite{goffman_nature, Watts_PNAS,Bart_CJS,feynmanSIR}.  Ideas, like diseases, have been shown to spread infectiously between people before eventually dying out, and have been successfully described with epidemiological models. Again, this follows intuitively, as ideas are spread through communicative contact between different people who share ideas with each other. Idea manifesters ultimately lose interest with the idea and no longer manifest the idea, which can be thought of as the gain of ``immunity" to the idea.

The epidemiological models presented in this study are used to analyze publicly available Google search query data for different OSNs, which can be obtained from Google's ``Google Trends" service \cite{googletrends}. Google search query data has been used in a range of studies, including the monitoring of disease outbreak \cite{Ginsberg}, economic forecasting \cite{Choi}, and the prediction of financial trading behavior \cite{Preis}. The work in Ref. \cite{Ginsberg} is particularly interesting, as the authors are able to correlate search query frequencies of flu-related terms to numbers of flu cases from CDC data, which is the basis of Google's ``Google Flu Trends" service \cite{FluTrends}. The results of Ref. \cite{Ginsberg} show that search query data can be used as a proxy for a tangible quantity (flu cases), with the advantage that search query data is quicker and easier to obtain than the actual aggregated reported flu case data\cite{Carneiro}. In this work we similarly use Google search query data as a proxy for another quantity of interest, OSN usership, for which data is difficult to obtain.

% You may title this section "Methods" or "Models". 
% "Models" is not a valid title for PLoS ONE authors. However, PLoS ONE
% authors may use "Analysis" 
\section*{Model}
\subsection*{Traditional SIR model}

The classical SIR model for the spread of infections disease is presented in Equations \ref{SIR_s}-\ref{SIR_r} \cite{Bailey1987}. The variable names are summarized in Table \ref{parameterTable}, which compares the epidemiological interpretation of the variables to the equivalent OSN dynamical interpretation of the variables. In short, the SIR model is a collection of three ordinary differential equations that govern the rates at which three compartments of the entire population $N$ (susceptible $S$, infected $I$, and recovered $R$) evolve during a disease outbreak, where the superscripted dot indicates a time derivative. Equations \ref{SIR_s}-\ref{SIR_r} sum to 0 when added together, revealing an underlying assumption that the population remains constant during the outbreak. Mathematically, $S+I+R=N$, independent of time. This assumption is valid for disease outbreaks that are short lived compared to the lifespan of the population members.

\begin{subequations}
\begin{eqnarray}
\dot S=-\frac{\beta IS}{N} \label{SIR_s}\\
\dot I=\frac{\beta IS}{N} - \gamma I \label{SIR_i}\\
\dot R=\gamma I\label{SIR_r}
\end{eqnarray}
\end{subequations}

Equation \ref{SIR_s} shows that the rate at which the susceptible population becomes infected is proportionate to the infection rate $\beta$, the fraction of infected population $\frac{I}{N}$, and the susceptible population $S$. This indicates that the disease is transferred through interaction between the susceptible and infected populations, with an infection rate of $\beta$. Equation \ref{SIR_r} shows that the rate at which the infected population recovers is proportionate to $I$ and the recovery rate $\gamma$, but does not require any sort of interaction between population compartments as was necessary for the spread of infection.

There is no analytical solution to the SIR model, but the set of ODE’s can be solved numerically with the addition of initial conditions for each of the population compartments: $S(0)=S_0$, $I(0)=I_0$ and $R(0)=R_0$. $S_0$ is the initial compartment of the population that is susceptible to contracting the disease, and generally represents the majority of the population at early times. $I_{0}$ represents the size of the initial outbreak, which is generally much smaller than the total population. $I_{0}$ must be non-zero for an infection to begin to spread as given by equation \ref{SIR_i}. Note that for the purposes of fitting the SIR model to a data set, if no initial parameters are known, than $R_0$ can be arbitrarily set to 0. This is because the magnitude of $R$ does not affect any of the dynamics other than contributing to the size of $N$. However, since terms containing $\beta$ are always divided by $N$, $R_0$ can be set arbitrarily and $\beta$ is then taken as a fitting parameter.

 One interesting feature of the model is the immunization criterion

\begin{equation}
\frac{S_{0} }{N} < \frac{\gamma}{\beta}\label{SIRimmune}
\end{equation}

which is derived by setting the right hand side of Equation \ref{SIR_i} less than or equal to 0 and rearranging. If Equation \ref{SIRimmune} is satisfied, then $\dot I$ is always less than or equal to 0, meaning that $I$ can never increase. This is the basis for immunization campaigns. In the context of OSNs, where OSN users are analogous to the infected population compartment, this immunization criterion represents the conditions under which a OSN will strictly decline.

The SIR model can be applied to OSNs by drawing the correct OSN analogues to the SIR model parameters. When applying the SIR model to OSNs, the susceptible population compartment is equivalent to all users that could potentially join the OSN. The infected population compartment is analogous to OSN users: potential OSN users are susceptible to joining the OSN through ``infection" by contact with a current OSN users. Finally, the recovered population compartment is analogous to the population of people who are opposed to joining the OSN. In the case of an OSN, this could be comprised of people who have left the OSN with no intention of returning or people who resist joining the OSN in the first place. A summary of all the OSN analogues to the epidemiological model parameters is provided in Table \ref{parameterTable}.

\subsection*{Infectious recovery SIR model} \label{sec_irSIR}

While the traditional SIR model captures the infectious uptake dynamics of OSNs, the assumption of a characteristic recovery rate in the traditional SIR model is of doubtful validity for modeling the abandonment of OSNs \cite{Wu2013}. Contrary to the SIR model's assumption of a characteristic recovery rate for diseases, OSN users do not join an OSN expecting to leave after a predetermined amount of time. Instead, every user that joins the network expects to stay indefinitely, but ultimately loses interest as their peers begin to lose interest. Thus a user that joins early on is expected to stay on the network longer than a user that joins later. Eventually, users begin to leave and “recovery” spreads infectiously as users begin to lose interest in the social network. The notion of infectious abandonment is supported by work analyzing user churn in mobile networks which show that users are more likely to leave the network if their contacts have left\cite{Dasgupta2008}. Therefore it is necessary to modify the traditional SIR model to include infectious recovery dynamics, which intuitively provide a better description of OSN abandonment.

To incorporate infectious recovery dynamics, the recovery rate must be modified to be proportionate to the recovered population fraction $\frac{R}{N}$, similar to how the spread of infection among the susceptible population in the SIR model is proportionate to the infected population fraction $\frac{I}{N}$. This is achieved by multiplying the $\gamma I$ terms in Equations \ref{SIR_i} and \ref{SIR_r} by $\frac{R}{N}$ to give Equations \ref{irSIR_i} and \ref{irSIR_r}. The constant $\gamma$ is replaced with a new constant $\nu$ which now represents an infectious recovery rate with the same units. 

\begin{subequations}
\begin{eqnarray}
\dot S=-\frac{\beta IS}{N} \label{irSIR_s}\\
\dot I=\frac{\beta IS}{N} - \frac{\nu IR}{N} \label{irSIR_i}\\
\dot R=\frac{\nu IR}{N} \label{irSIR_r}
\end{eqnarray}
\end{subequations}

In modifying Equation \ref{irSIR_i} and \ref{irSIR_r}, $R_{0}$ is no longer an arbitrary parameter as it was in the SIR model, because the magnitude of $R$ is now important to the recovery dynamics. This results in an additional fitting parameter when applying the irSIR model to the data. Similar to the requirement of an initial infected population in the traditional SIR model, the irSIR model necessarily requires a small initial recovered population. If $R_{0}$ were set to 0, none of the infected population would be able to recover since recovery now requires contact with a recovered individual. Mathematically, Equation \ref{irSIR_r} would be zero for all time, and $I$ would grow at the expense of $S$ until the entire population is infected. In the context of OSN dynamics, $R_{0}$ can be thought of as the first users to leave the OSN or a compartment of the population that resists joining the OSN altogether. This small initial compartment of OSN resistors is ultimately responsible for the abandonment of the OSN.

An immunization criterion can be derived for the irSIR model similar to Equation \ref{SIRimmune}

\begin{equation}
\frac{S_{0}}{R_{0}}<\frac{\gamma}{\beta}\label{irSIRimmune}
\end{equation}

Similar to the implications of Equation \ref{SIRimmune} in the traditional SIR model, if this criterion is satisfied then $I$ can never increase. One might also wonder if an analogous criterion for ``immunization" against recovery can be derived for the irSIR model, since the recovery and infection dynamics are now the same. In the context of OSNs, immunization against recovery would ensure that the OSN always grows in time. Looking at Equation \ref{irSIR_r}, one can see that unless $R_{0}=0$, $\dot R$ is always positive and $R$ must always increase in time. The implication is that there is no way to ``immunize" the infected population compartment against infectious recovery in the irSIR model. In the context of the irSIR model, all OSNs are expected to eventually decline.

\section*{Methods}
\subsection*{Google Trends Search Query Data}

To test the theory of disease-like adoption dynamics of OSNs, we use publicly available historical Google search query data as a proxy for OSN usership, as has been done in previous work \cite{Garcia2013}. The public nature of this data is an important feature of the approach used in this paper, as historical OSN user activity data is typically proprietary and difficult to obtain \cite{Torkjazi2009}. The Google search query data is obtained from Google's ``Google Trends" service and reports the relative number of Google search queries for a given search term \cite{googletrends}. The use of search query data for the study of OSN adoption is advantageous compared to using registration or membership data in that search query data provides a measure of the level of web traffic for a given OSN. Web traffic is arguably the best metric for OSN health in that it represents a measure of user activity or interest within the network. For example, inactive members do not contribute to a social network and would not be counted using search data, but would be counted using other metrics such as registered account data. Thus OSN usership as measured by search query data can be thought of as representing a less tangible, albeit more meaningful, metric of user activity or interest.

The search query data obtained from Google Trends is presented in arbitrary units of weekly search query frequency which are normalized such that the maximum data point in the selected time period corresponds to a value of “100.” The use of weekly search query data is advantageous over daily data because it eliminates periodic variation due to day of week. For example, search queries for ``Facebook" tend to peak during the weekend. The resolution of this search query data is 1 unit. To increase the resolution of the search query data for model fitting, shorter time periods of data are stitched together. For example, consider two consecutive sets of weekly search query data chosen such that the last week of the first data set and the first week of the second data set are the same. If that week has a value of 90 in the first data set and 100 in the second, then the two data sets can be stitched together by multiplying the second data set by a factor of 90/100 so that they are now on the same scale. This process can be repeated with data spanning the entire time range of interest and the resulting data set is subsequently normalized such that 100 corresponds to the maximum data point. The ultimate result is a data set with resolution many times higher than the original 1 unit.

\begin{figure}[!ht]
  \begin{center}
  \includegraphics[width=4in]{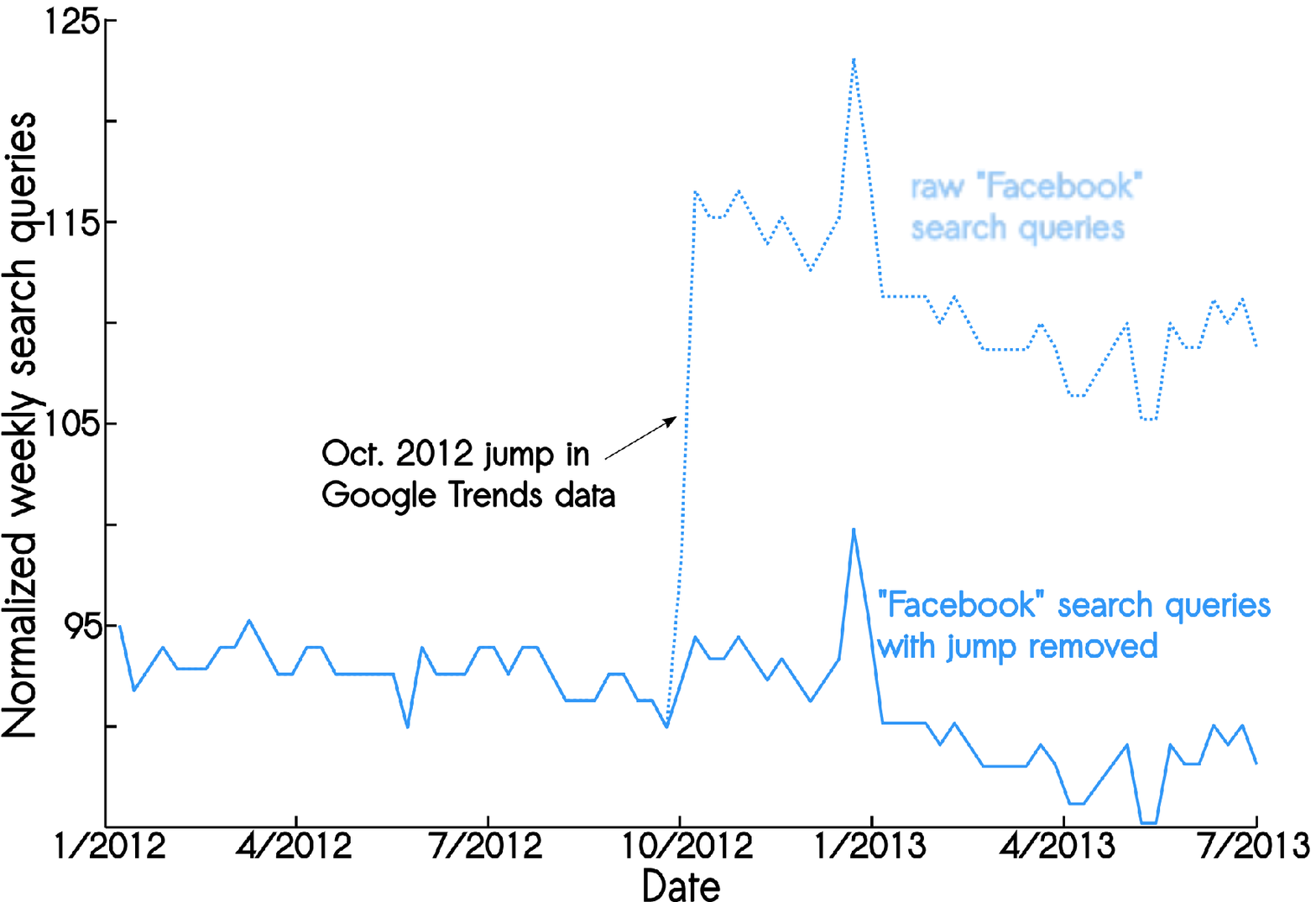}
  \end{center}  
  \caption{\textbf{Google search query data for ``Facebook" between January 2012 and July 2013 before and after removal of the artifactual October 2012 jump in search queries.} Both data sets are scaled such that 100 corresponds to the maximum weekly Google search queries for the set with the jump removed over the plotted time period.}
  \label{trendsJump}
  \end{figure}

The Google search query data for the term “Facebook” shows a large jump in search queries during early October 2012 as shown in Figure \ref{trendsJump}. This jump is assumed to be artifactual, as it is unlikely that search activity for Facebook suddenly jumped by 20\% in under a week and remained consistently at this elevated level in the weeks following. The authors believe this artifact is due to the October 5, 2012 upgrade of Google’s search algorithm to Penguin 3 \cite{penguinUpdate}, which coincides with the reported jump in Facebook search queries. A similar jump, albeit smaller in magnitude, can also be observed in the Google search query data for other search terms such as ``Twitter." To the authors’ knowledge, there is no publicly available explanation for how this update should affect the Google search query data reported by Google Trends. This jump is removed for the purposes of data analysis in this paper by multiplying the search query data occurring after the week ending October 6, 2012 by a factor of 0.804. A multiplicative 
factor is used instead of subtraction of a constant because it is assumed to be more likely that an artifactual change in search query reporting would affect all search queries equally, as opposed to simply removing a constant amount of searches queries each week. 

\begin{figure}[!ht]
  \begin{center}
  \includegraphics[width=3in]{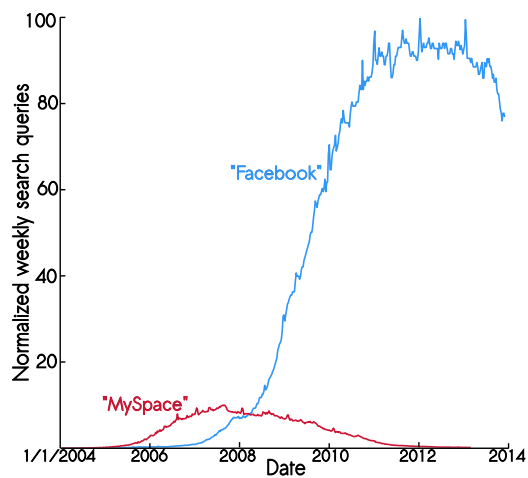}
  \end{center}
  \caption{\textbf{Search query data for ``Facebook" and ``MySpace" obtained from Google Trends overlaid on top of each other.} The data for both curves are scaled such that 100 corresponds to the maximum weekly Google search queries for "Facebook" over the plotted time period. Search queries for ``MySpace" peak at 10\% of the maximum weekly search queries for ``Facebook" in this time period.}
  \label{Gtrends}
  \end{figure}

The Google data for search query ``MySpace" and search query ``Facebook" are presented in Figure \ref{Gtrends}. In Figure \ref{Gtrends} the MySpace data is normalized to the same scale as the Facebook data in order to compare the relative number of search queries for both OSNs. Figure \ref{Gtrends} shows a number of important features that qualitatively validate the use of search query data as a proxy for OSN usership. The shape of the MySpace curve in Figure \ref{Gtrends} is in qualitative agreement with the timeline for the rise and fall of MySpace discussed in Section \ref{sec_Intro} \cite{myspaceFall}, exhibiting a rise in 2005, a peak between 2007 and 2008, and a steady decline between 2009 and 2011. The intersection of the MySpace and Facebook curves occurs near April, 2008, the month in which Facebook is reported to have overtaken MySpace in terms of global web traffic \cite{FBovertakes}. The Facebook curve in Figure \ref{Gtrends} shows a decline in search activity starting in 2013, which corroborates reports that the OSN started losing some of its younger user base in 2013 \cite{FBtranscript,DM1}.

\subsection*{Search Query Data Curve Fitting}

The Google search query data for the OSN is assumed to be representative of the magnitude of the infected population compartment of the previously discussed epidemiological models. Recall that the infected compartment in the context of OSNs is the active user base of the OSN. The epidemiological models are used to analyze the search query data by curve fitting the infected $I(t)$ population curve generated by the model to the search query data. The curve fitting is carried out using the MATLAB computational software package \cite{matlab} using the following approach to determine the best fit curve.

To find the best fit curve, an $I(t)$ curve must be generated and then compared to the search query data set. This is achieved by choosing initial input parameters $S_0$, $I_0$, $R_0$, $\beta$, and $\nu$, and then solving the system of differential equations to determine the corresponding $I(t)$. The sum of squared error (SSE) between the weekly search query data points and the generated $I(t)$ curve is then calculated. The best fit curve is defined as the $I(t)$ curve that minimizes the calculated SSE. The best fit curve is determined using a built in MATLAB function, {\fontfamily{pcr}\selectfont  fminsearch}, which uses a Nelder-Mead simplex method \cite{fminsearch} to find determine the set of 5 input parameters corresponding to the best fit $I(t)$ curve. This optimization scheme uses a Dormand-Prince (Runge-Kutta 4,5) method \cite{ode45} to compute $S(t)$, $I(t)$, and $R(t)$. Once the best fit curve is determined, the corresponding input parameters are saved so that the best fit curve 
can now be generated by re-solving the system of differential equations using the best fit parameters. 

\section*{Results and discussion}

\subsection*{MySpace}

\begin{figure}[!ht]
  \begin{center}
  \includegraphics[width=4in]{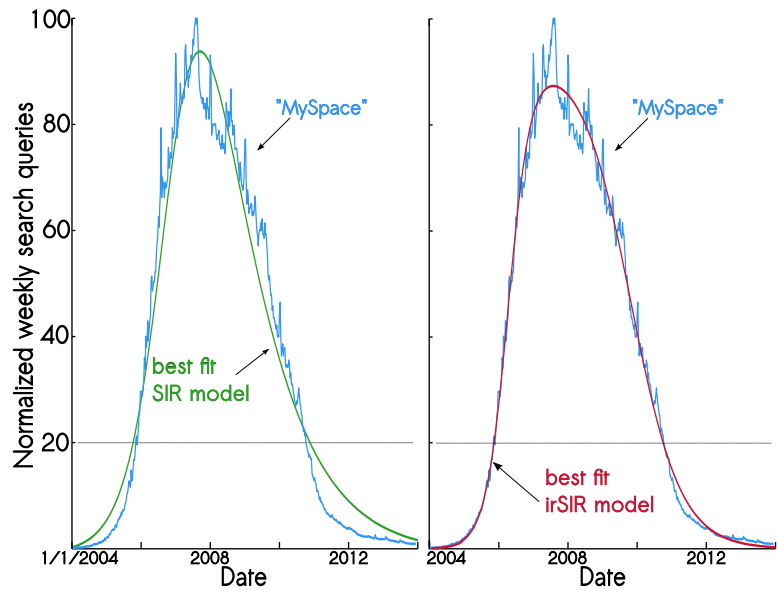}
  \end{center}
  \caption{ \textbf{Data for search query ``Myspace" with best fit (a) SIR and (b) irSIR models overlaid.} The search query data are normalized such that the maximum data point corresponds to a value of 100.}
  \label{myspaceFit}
  \end{figure}

To validate the use of epidemiological models in OSN adoption dynamics, we first consider the case of MySpace. MySpace is a particularly useful case study for model validation because it represents one of the largest OSNs in history to exhibit the full life cycle of an OSN, from rise to fall. MySpace also has the added advantage of having its entire lifespan occur within the range of search query data available from Google Trends, which is only available after January 2004. The data for search query “MySpace” is shown in Figure \ref{myspaceFit} as the solid blue line. This is the same search query data that is presented in Figure \ref{Gtrends}, but in Figure \ref{myspaceFit} the data is normalized such that the maximum value of weekly Google search queries for ``MySpace" corresponds to a value of 100. 

To fit the Myspace data, we first apply the SIR model. As shown in Figure \ref{myspaceFit}a, the best fit SIR model provides a somewhat qualitative description of the search query data, although it clearly fails to fit near the shoulders of the curve. The relatively poor fit is expected given that the constant recovery rate assumption in the SIR model does not intuitively describe the adoption and abandonment dynamics of OSNs. Rather, as discussed in detail in Section \ref{sec_irSIR}, ``recovery" spreads infectiously: users begin to leave the OSN after their peers have left the OSN.

To capture the intuitive infectious recovery dynamics for OSN abandonment, the irSIR model is applied to the Myspace data as shown in Fig. \ref{myspaceFit}b. The irSIR model provides a significantly better description of the search query data, hugging the shoulders of the curve more tightly and resulting in a 75\% drop in SSE as seen in Table \ref{fitTable}. The quality of fit of the irSIR model lends validity to the notion of infectious user abandonment of OSNs. 

\subsection*{Facebook}
Having validated the application of the irSIR model to the adoption dynamics of OSNs, we now apply the irSIR model to the Google search query data for Facebook. Facebook presents an interesting case study as it is the largest OSN in history. Moreover, the search query data suggests that Facebook has already reached the peak of its popularity and has entered a decline phase, as evidenced by the downward trend in search frequency after 2012. The presence of a year-long decline portion of the curve allows for determination of best fit $R_0$ and $\nu$ parameters, both of which are sensitive to the decline dynamics. Given the early stage of Facebook's decline at the time of writing, the best fit irSIR model can be used not only to provide an explanation for the observed curve, but also to forecast how the OSN will decline in the future according to the assumptions of the irSIR model. 

Applying the irSIR model to Facebook shows a high quality fit for the available search query data over the time period of January 2004 to the last reported data point at the time of writing. Extrapolating the best fit into the future shows that Facebook is expected to undergo rapid decline in the upcoming years, shrinking to 20\% of its maximum size by December 2014. We will use 20\% as an arbitrary definition for end of life so that we can define a so-called ``20\% date" as the OSN's end of life date.

\begin{figure}[!ht]
  \begin{center}
  \includegraphics[width=3in]{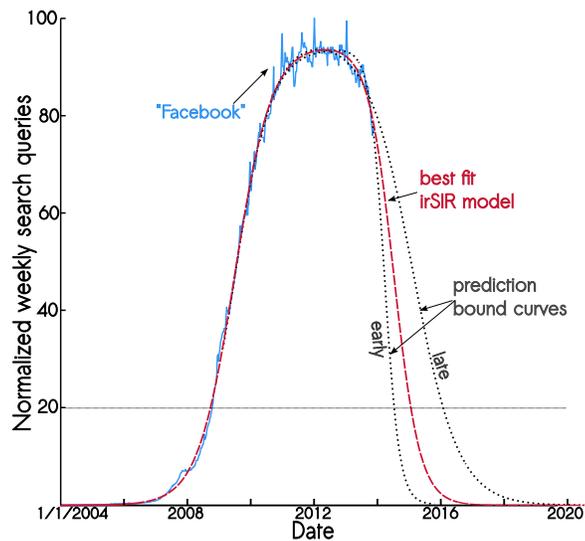}
  \end{center}
  \caption{\textbf{Data for search query ``Facebook" with best fit irSIR model overlaid.} Prediction bounds corresponding to the earliest and latest decline times with a 15\%  increase in SSE over the best fit are overlaid. The data are scaled such that the maximum weekly search queries for ``Facebook," corresponds to a value of 100.}
  \label{facebookFit}
  \end{figure}

To illustrate the strength of the prediction, upper and lower model prediction bounds for the abandonment of Facebook are overlaid alongside the best fit irSIR model in Figure \ref{facebookFit}. The corresponding fitting parameter values are tabulated in Table \ref{fitTable}. The prediction bounds plotted in Figure \ref{facebookFit} represent the two curves with the earliest and latest 20\% dates that exhibit less than or equal to 15\% higher SSE than the true best fit curve. Thus all possible irSIR model solutions with an increase in SSE of less than a 15\% over the best fit are bounded between the plotted prediction bounds. The presentation of a bounded prediction region emphasizes that the plotted best fit curve should be thought of as the best forecast given the available data.

The differences between the 15\% SSE prediction bound curves and the best fit curve are mainly determined by the $R_0$ and $\nu$ parameters, which follows from the importance of these parameters in describing the abandonment of the OSN in the irSIR model. The early  prediction bound curve has a higher $\nu$ parameter than the best fit model, which is responsible for the higher rate of decline in the early prediction bound. To compensate for the rapid decline, $R_0$ is decreased such that the onset of the decline is later and the error bound coincides with declining data occurring after 2012. The opposite is true for the late prediction bound curve, which has a lower $\nu$ and higher $R_0$ than the best fit. This results in a curve that declines slower than the best fit, but also begins declining earlier such that it coincides with the 2013 data. 

It is also important to understand the nature of the asymmetry of the prediction bounds. The difference in 20\% dates between the late bound and the best fit model is much larger than the difference in 20\% dates between the best fit model and the early bound. Since the early and late prediction bound curves can be thought of as equally probable future trajectories for Facebook, it follows that Facebook is more likely to deviate from the best fit curve on the side of slower decline. The reason for this asymmetry is that the decline is entirely dictated by data occurring in after 2012. If the post-2012 data is ignored, a solution in which Facebook continues indefinitely at a constant size is possible with a finite SSE determined mainly by deviation from the post-2012 data. This implies that there exists an infinite range of possible slower declining solutions than the best fit, all with SSE values bounded primarily by deviation from the post-2012 data. Conversely, solutions declining faster than the best fit are bounded by a lower limit set by the date of the last reported data point in the data set.

\section*{Conclusions}
In this paper we have applied a modified epidemiological model to describe the adoption and abandonment dynamics of user activity of online social networks. Using publicly available Google data for search query “Myspace” as a case study, we showed that the traditional SIR model for modeling disease dynamics provides a poor description of the data. A 75\% decrease in SSE is achieved by modifying the traditional SIR model to incorporate infectious recovery dynamics, which is a better description of OSN dynamics. Having validated the irSIR model of OSN dynamics on Google data for search query “Myspace,” we then applied the model to the Google data for search query “Facebook.” Extrapolating the best fit model into the future suggests that Facebook will undergo a rapid decline in the coming years, losing 80\% of its peak user base between 2015 and 2017.

% Do NOT remove this, even if you are not including acknowledgments
\section*{Acknowledgments}
This publication is the culmination of talk originally presented at the 2012 Princeton Research Symposium titled "The social network disease: epidemiology and the demise of Facebook." The authors acknowledge PRS for its provision of an intellectually stimulating forum for the discussion of research, without which this work would not have been conceived. The authors acknowledge the Princeton Graduate School, specifically Dean William B. Russel, for continued support of PRS and help in meeting the costs of publication. In addition, the authors acknowledge Professor Craig B. Arnold for fruitful golf discussion. 

%\section*{References}
% The bibtex filename
\bibliography{references3}

\section*{Figure Legends}
%\begin{figure}[!ht]
%\begin{center}
%%\includegraphics[width=4in]{figure_name.2.eps}
%\end{center}
%\caption{
%{\bf Bold the first sentence.}  Rest of figure 2  caption.  Caption 
%should be left justified, as specified by the options to the caption 
%package.
%}
%\label{Figure_label}
%\end{figure}

% ******************************
% ******************************

% ******************************
% ******************************

% ******************************
% ******************************

% ******************************
% ******************************

\section*{Tables}
%\begin{table}[!ht]
%\caption{
%\bf{Table title}}
%\begin{tabular}{|c|c|c|}
%table information
%\end{tabular}
%\begin{flushleft}Table caption
%\end{flushleft}
%\label{tab:label}
% \end{table}

\begin{table}[!ht]
\caption{
\bf{Epidemiological model parameter definitions}}
\begin{tabular} {|c|c|c|c|}
\hline
Symbol &  Units & Disease Model Parameter & Equivalent OSN Model Parameter \\ \hline
S & People & Susceptible & Potential OSN users \\ \hline
I & People & Infected & OSN users \\ \hline
R & People & Recovered/Immune & Population opposed to OSN use \\ \hline
$\beta$ & Time$^{-1}$ & Infection rate & Rate at which potential users join OSN \\ \hline
$\gamma$ & Time$^{-1}$ & recovery rate & - \\ \hline
$\nu$ & Time$^{-1}$ & - & OSN abandonment rate \\
\hline
\end{tabular}
\begin{flushleft}Summary of parameters in the epidemiological models and their equivalent OSN analogues.
\end{flushleft}
\label{parameterTable}
\end{table}

\begin{table}[!ht]
\caption{
\bf{Epidemiological model fitting parameters}}
\begin{tabular} {|c|c|c|c|c|c|c|c|c|}
\hline
Fit &  $\beta$ & $\nu$ & $\frac{S_0}{N}$ & $\frac{I_0}{N}$ & $\frac{R_0}{N}$ & $N$ & $SSE$ & $20\%$ date \\ \hline

MySpace, SIR & $4.92 \cdot 10^{-2}$ & 5.39 ($\gamma$) & 0.996 & $4.1 \cdot 10^{-3}$ & 0 (fixed)& 322 & $1.67\cdot10^{4}$ & 12/2010 \\ \hline %n=517

MySpace, irSIR & $5.98 \cdot 10^{-2}$ & $2.68 \cdot 10^{-2}$ & 0.992 & $9.49 \cdot 10^{-4}$ & $7.19 \cdot 10^{-3}$ & 92.94 & $4.33 \cdot 10^{3}$ & 11/2010 \\ \hline %n=517

Facebook, best & $3.36 \cdot 10^{-2}$ & $4.98 \cdot 10^{-2}$ & 1.00 & $6.43 \cdot 10^{-5}$ & $2.35 \cdot 10^{-6}$ & 94.5 & $1.34 \cdot 10^{3}$ & 1/2015 \\ \hline %n=516 %t20=wk571

Facebook, early & $3.43 \cdot 10^{-2}$ & $8.23 \cdot 10^{-2}$ & 1.00 & $5.47 \cdot 10^{-5}$ & $1.04\cdot 10^{-9}$ & 93.61 & $1.53 \cdot 10^{3}$ & 07/2014 \\ \hline %n=516 %t20=wk534

Facebook, late & $3.27 \cdot 10^{-2}$ & $2.71 \cdot 10^{-2}$ & 1.00 & $8.09 \cdot 10^{-5}$ & $4.10 \cdot 10^{-4}$ & 97.0 & $1.54 \cdot 10^{3}$ & 02/2016 \\ \hline %n=516 %t20=wk639

\end{tabular}
\begin{flushleft}Table showing fitting parameter values from fitting of search query data with epidemiological models.
\end{flushleft}
\label{fitTable}
\end{table}

\end{document}